# Biologically inspired graphene-chlorophyll phototransistors with high gain

*Shao-Yu Chen[1], Yi-Ying Lu[1,2], Fu-Yu Shih[1,3], Po-Hsun Ho[4], Yang-Fang Chen[3],*

*Chun-Wei Chen[4], Yit-Tsong Chen[1,2], and Wei-Hua Wang[1*]*

[1]Institute of Atomic and Molecular Sciences, Academia Sinica, No. 1, Roosevelt Rd., Sec. 4, Taipei 106, Taiwan

[2]Department of Chemistry, National Taiwan University, No. 1, Roosevelt Rd., Sec. 4, Taipei 106, Taiwan

[3]Department of Physics, National Taiwan University, No. 1, Roosevelt Rd., Sec. 4, Taipei 106, Taiwan

[4]Department of Materials Science and Engineering, National Taiwan University, No. 1, Roosevelt Rd., Sec. 4, Taipei 106, Taiwan

[*]Corresponding Author.

Tel: +886-2-2366-8208, Fax: +886-2-2362-0200. E-mail address: wwang@sinica.edu.tw (W.-H. Wang)

**Abstract**

We present prominent photoresponse of bio-inspired graphene-based phototransistors sensitized with chlorophyll molecules. The hybrid graphene-chlorophyll phototransistors exhibit a high gain of $10^6$ electrons per photon and a high responsivity of $10^6$ A/W, which can be attributed to the integration of high-mobility graphene and the photosensitive chlorophyll molecules. The charge transfer at interface and the photogating effect in the chlorophyll layer can account for the observed photoresponse of the hybrid devices, which is confirmed by the back-gate-tunable photocurrent as well as the thickness and time dependent studies of the photoresponse. The demonstration of the graphene-chlorophyll phototransistors with high gain envisions a viable method to employ biomaterials for graphene-based optoelectronics.

**Manuscript text**

### 1. Introduction

Graphene-based optoelectronic devices have been attracting great attention[1-3] and have promise for future applications, such as solar cells, touch screens, and photodetectors.[4-6] The remarkable optical properties of graphene, including its linear optical absorption[7, 8], tunable band-gap,[9] and intrinsic photocurrent[10-13] have been demonstrated and can be integrated with other distinctively electronic and mechanical properties of graphene for new functionalities. Pristine graphene suffers from weak absorption as low as 2.3%, which fundamentally limits its applicability in photodetection. Consequently, various approaches, including plasmonic effects,[14, 15] photothermoelectric effects,[16] and sensitization by quantum dots[6, 17, 18] or bulk semiconductors,[19] have been employed to enhance the interaction between graphene and photons. Here, we utilize chlorophyll as light absorbing materials for graphene phototransistors and investigate the optoelectronic performance of the hybrid graphene-chlorophyll devices. Chlorophyll is an organic semiconductor and is very efficient in light absorption. Moreover, chlorophyll, as a biomaterial, is remarkably stable[20] and abundant. Given the crucial role of chlorophyll in the photosynthesis process, studies of chlorophyll photosensitization may benefit future applications, such as organic photovoltaics,[21, 22] optical sensors,[23] and artificial photosynthesis.[24] Additionally, since there are many types of chlorophyll-related molecules, it is possible to engineer the energy level alignment such that the charge separation and transfer are optimized in the hybrid graphene-organic molecules systems.

In this paper, we present strong photoresponse observed in bio-inspired graphene-chlorophyll devices. A pristine graphene/chlorophyll interface condition is achieved by employing resist-free fabrication with *in-situ* and noncovalent functionalization. A high gain and a high responsivity of the hybrid graphene-chlorophyll devices are demonstrated and attributed to the integration of the high-mobility graphene and photosensitive chlorophyll molecules. Effective photogating effect originated from charge separation/transfer can account for the observed large photoresponse, which is confirmed with the back-gate-tunable photocurrent as well as the thickness and time dependent studies of the photoresponse.

## 2. Device Fabrication

A detailed description of the device fabrication and functionalization of chlorophyll *a* (Chl*a*) can be found in the supporting information (S1 and S2). We note that we have adapted careful procedures to ensure a pristine graphene/Chl*a* interface. First, monolayer graphene flakes were mechanically exfoliated onto the octadecyltricholorosilanes (OTS)-functionalized substrates, and electrical contacts were fabricated by resist-free approach to achieve a high mobility and a pristine graphene surface.[25] Second, we performed laser annealing with excitation adjacent to the graphene devices, which greatly reduced adsorbed molecules on the graphene surface (see supporting information S1). Third, we employed controlled drop-casting[26] of Chl*a* on-site of the cryostat with a continuous nitrogen gas flow. Therefore, the cleanness of the graphene surface was preserved, which resulted in an optimum contact condition between the Chl*a* molecules and graphene. By following these careful procedures of device fabrication, we achieved reproducible results on both sensitization and optoelectronic properties of the phototransistors.

## 3. Results and Discussion

To confirm the effectiveness of surface functionalization, we performed Raman spectroscopy and conductivity measurement on the graphene samples. Figure 1a shows the Raman spectrum of pristine graphene (sample A) with an intensity ratio of 2D/G band of approximately 6 and a negligible D-band, which indicates the high quality of the pristine graphene. After Chl*a* functionalization, the Raman spectrum reveals extra peaks in the range of 1000-1600 $cm^{-1}$, which generally agrees with the resonance spectra of the Chl*a* film.[27] Moreover, we characterized the devices electrically by measuring the conductance ($G$) as a function of the gate voltage ($V_G$). Figure 1b shows the $G-V_G$ curves of sample A before and after Chl*a* functionalization, which reveals the n-type doping characteristics as $V_G$ at the charge neutrality point ($V_{CNP}$) shifts from $-5$ V to $-40$ V. The n-type doping of graphene suggests that the electrons transfer from Chl*a* to graphene after the functionalization.

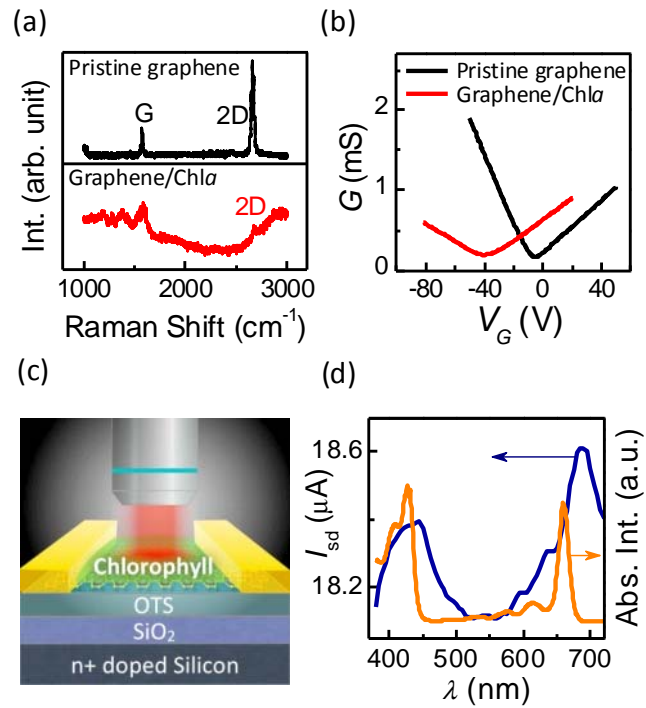

**Figure 1.** (a) Raman spectra before (black curve) and after (red curve) Chl*a* functionalization. A group of Raman peaks in the range of 1000 ~ 1600 cm$^{-1}$ are observed and are attributed to the Chl*a* film. (b) $G-V_G$ curves before (black curve) and after (red curve) Chl*a* functionalization ($V_{SD}=100$ mV). A reduction of mobility from 6500 to 1600 $cm^2/V\cdot s$ for hole conduction and from 2700 to 1850 $cm^2/V\cdot s$ for electron conduction was observed. (c) A schematic of the graphene-Chl*a* device on an OTS-functionalized substrate under illumination. (d) Photocurrent action spectrum of sample B (blue curve). The two-peak structure observed at about 430 nm and 690 nm generally corresponds to the absorption spectrum of the Chl*a* solution (orange curve).

The charge transfer can be attributed to the energy level alignment of graphene and Chl*a*, as shown in Figure 2b.[28] Because the Fermi level of Chl*a* ($E_{F,Chla} \approx -4.1$ eV) is higher than that of graphene ($E_{F,Gra} \approx -4.6$ eV)[29, 30], $E_{F,Gra}$ is raised when the two materials are brought into contact and the electrons tend to transfer from Chl*a* to graphene. The charge transfer consequently causes both n-type doping in graphene and a positively charged region near the interface at Chl*a* side.[31] Moreover, the positively charged region results in a built-in electric field at the interface that points toward graphene.[32-34] It is noted that the n-type doping of graphene is consistently observed in all 5 samples that we fabricated, with $\Delta V_{CNP}$ ranging from 30 to 35 V. Therefore, both Raman spectroscopy and conductivity measurements of the samples suggest the effective functionalization of Chl*a* molecules on graphene.

We then measured the spectral photocurrent to further examine the role of Chl*a* molecules in the hybrid system. Figure 1c illustrates a schematic of a graphene-Chl*a* phototransistor under illumination. In Figure 1d, we plot the photocurrent action spectrum of a test graphene-Chl*a* device (sample B) that shows the Soret band (430 nm) and the $Q_y$ band (690 nm) with three vibronic bands. The shape of the photocurrent action spectrum of sample B generally agrees with the shape of the absorption spectrum of Chl*a* in the solution phase shown in the same figure. The peaks in the action spectrum are red-shifted as compared with those in the absorption spectrum, which has been previously reported in ordered structure of Chl*a*.[35, 36] The redshift can be attributed to the stacking of the porphyrin rings of Chl*a* when it aggregates, leading to a lowering in excitation energy.[37] Hence, the above result suggests that the photoinduced excitation in the Chl*a* film initiates the charge transfer, which generates the photocurrent in the graphene channel.

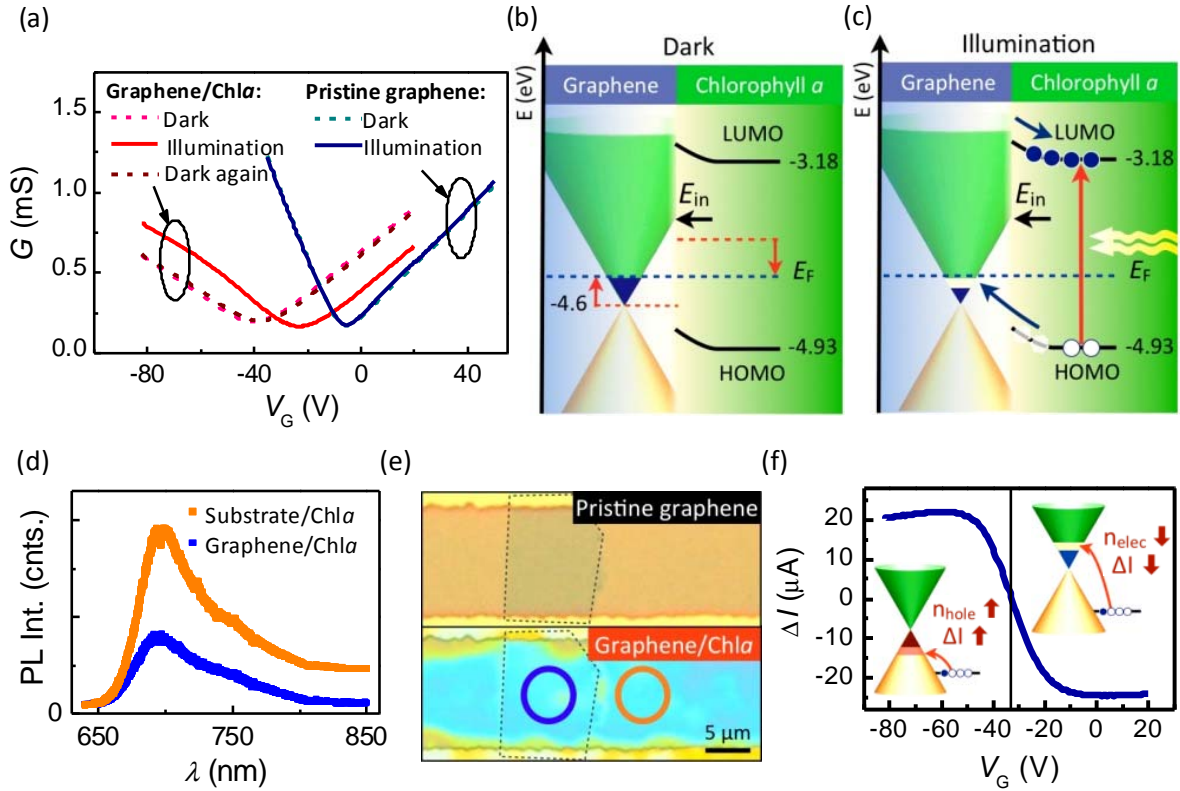

**Figure 2.** (a) Comparisons of the $G-V_G$ curves before and after Chl*a* functionalization at various stages ($V_{SD}=100$ mV). For sample A, $V_{CNP}$ shifts positively for ~17 V at an illumination power of 22 mW/cm$^2$. The excitation wavelength in all photoresponse measurements is 683 nm. The energy level alignment for (b) n-doping effect due to Chl*a* functionalization and (c) photodoping effect under illumination. (d) The PL spectra of graphene/Chl*a* (blue curve) and substrate/Chl*a* (orange curve) corresponding to the positions marked as blue and orange circles in Figure 2e. (e) The OM images of sample A before and after Chl*a* functionalization. (f) The photocurrent as a function of gate voltage, which shows a highly tunable capability. ($V_{SD}=100$ mV) Inset: The schematics that show the direction of hole transfer from Chl*a* to graphene for the hole and electron conduction regimes in graphene under illumination.

The large photoresponse due to Chl*a* sensitization is presented in Figure 2a, where we compare the $G-V_G$ curves of the pristine graphene and the graphene-Chl*a* device (sample A) at various stages. For pristine graphene, no significant photoresponse is observed with $V_G$ remaining at -5 V regardless of the illumination power. By contrast, sample A exhibits a large photodoping effect, with $V_{CNP}$ shifting substantially toward positive $V_G$ from -40 to -23 V. We note that the observed photoinduced doping in this study is larger than that in the graphene systems sensitized with quantum dots[18] and $C_{60}$,[38] suggesting the uniqueness of the graphene-Chl*a* system. The photoinduced carrier concentration in sample A can be estimated by the equation $n = \alpha \Delta V_{CNP} \approx 10^{12}$ cm$^{-2}$, where $\alpha$ is the carrier density capacitance of graphene. Furthermore, $V_{CNP}$ shifts back to -40 V after the laser irradiation, revealing a reversible photoresponse process.

Here, we propose a qualitative model to account for the observed photoresponse in our samples (Figure 2c). The hole carriers in Chl*a* are energetically favorable to transfer to graphene because the work function of graphene is smaller than the ionization potential of Chl*a* (-4.93 eV).[39] Moreover, this charge transfer of the hole carriers is assisted by the built-in electric field at the interface that points toward graphene. Accordingly, hole carriers are photoinduced in the Chl*a* film, and those hole carriers near the graphene/Chl*a* interface are swept into graphene, which leads to efficient charge separation and transfer. Conversely, the electron carriers tend to reside in the Chl*a* layer due to the existence of the built-in field. Therefore, the Chl*a* film becomes more negatively charged, which induces positive charge and p-type doping in graphene. The observed large photoinduced doping can be attributed to efficient charge separation/transfer at the interface and the strong photogating effect due to the proximity of the charged Chl*a* molecules to graphene.

Next, we present photoluminescence (PL) and back-gate dependent photocurrent measurements that further support the proposed model. We compare the PL of graphene-Chl*a* and Chl*a* on the substrate (Figure 2d), which reveals an apparent quenching of PL in Chl*a* on graphene. Because electron-hole pairs are generated in the Chl*a* molecules under illumination, the PL quenching suggest that charge transfer occurs between the Chl*a* film and graphene, which leads to reduction of radiative recombination.[40] Moreover, the back-gate dependence of the photoresponse (Figure 2f) is also consistent with the proposed model. We observed a strong carrier type and concentration dependence of the photocurrent ($\Delta I \equiv I_{light} - I_{dark}$), where $\Delta I$ is positive for hole conduction and negative for electron conduction. In the hole conduction regime ($V_G < -33$ V), carrier concentration increases because excess holes are induced by the photogating effect, which raises the conductance of graphene and a positive $\Delta I$. By contrast, the photoinduced hole carriers diminish the electron concentration, which leads to a decreased conductance and a negative $\Delta I$ in the electron conduction regime ($V_G > -33$ V). Therefore, the photocurrent of the device can be effectively tuned by back gate, and its polarity can also be reversed.

We can obtain the specifics of the devices from the power dependence of the photoresponse. Figures 3a and 3b show the power dependent evolution of resistance ($R$) vs. $V_G$ curves for the pristine graphene and sample A, respectively. Compared to relatively power independent $V_{CNP}$ of pristine graphene, sample A exhibits a monotonic shift in $V_{CNP}$ toward positive $V_G$. In Figure 3c, $\Delta V_{CNP}$ monotonically increases from 0 to 17 V with illumination power, which indicates an accumulative charge transfer between Chl*a* and graphene.[41] The photoinduced carrier concentration is $n \approx 10^{12}$ cm$^{-2}$ for $P = 100$ mW/cm$^2$. $\Delta V_{CNP}$ starts to level off at a high excitation power ($P > 1$ mW/cm$^2$), suggesting that the charge transfer begins to saturate. Regarding the transport characteristics, we observe that minimum conductivity ($\sigma_{min}$) decreased with an increasing excitation power as shown in Figure 3d. As discussed earlier, the functionalization of graphene results in a positively charged Chl*a* film, which causes charge inhomogeneity in graphene.[42, 43] The charge transfer in the device under illumination reduces the positive charging in Chl*a* film, which leads to the reduction of charge inhomogeneity in graphene and its minimum conductivity (supporting information S3). Therefore, the result suggests a trend toward the intrinsic state of graphene upon illumination. We note that this behavior is consistent with smaller doping concentration under excitation as can be seen in Figure 2a.

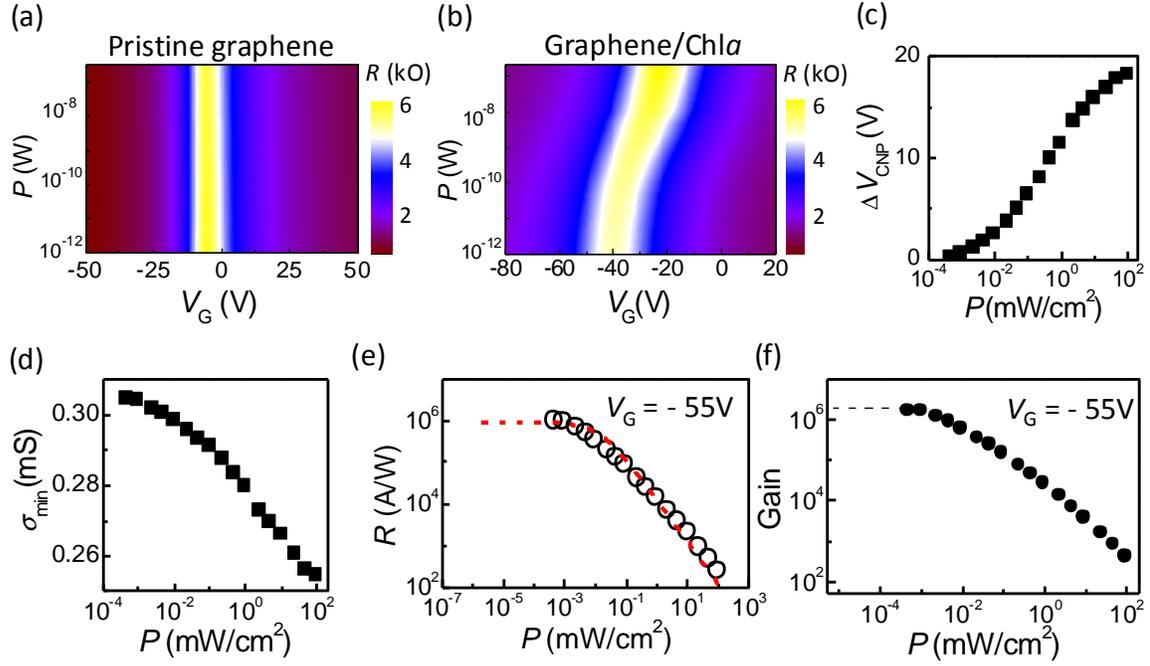

**Figure 3.** Power dependent evolution of the $R-V_G$ curves for (a) pristine graphene and (b) sample A, which shows that $V_{CNP}$ shifts monotonically toward a positive gate voltage with increasing power in the graphene/Chl$a$ device. $V_{SD}=100$ mV for all the measurements shown in this figure. (c) The shift of $V_{CNP}$ as a function of power density. (d) The minimum conductivity as a function of the excitation power. (e) The responsivity of the device is shown as a function of the illumination power and is as high as $1.1\times10^6$ A/W for an illumination power of 440 nW/cm$^2$ and $V_G=-55$ V. (f) The photoconductive gain as a function of the illumination power: the highest gain reaches $1.7\times10^6$ at power of 440 nW/cm$^2$ and $V_G=-55$ V.

Figure 3e shows the responsivity of the device as a function of the illumination power, which can be fitted by the equation $R=C_1/(C_2+P)$,[6] where $C_1$ and $C_2$ are fitting parameters. The built-in electric field, which causes the photoinduced doping in graphene, is reduced when more hole carriers are transferred to graphene under higher power (Figure 2c). This consequently results in decreasing responsivity at increasing illumination power. Notably, the responsivity reaches $10^6$ A/W for an illumination power of 440 nW/cm$^2$, and the device can detect a laser power as low as 500 fW. The responsivity in our device is comparable to that of graphene-based photodetectors previously reported.[17, 18] We also calculated the photoconductive gain as a function of illumination power (Figure 3f). The theoretical photoconductive gain $G_{th}$, as defined by the ratio of the change of the carriers in graphene to the change of the photons illuminated, can be written as

$$G_{th} = \frac{\Delta\sigma \cdot V_{SD}}{eA\phi_{photon}},$$

where $\Delta\sigma$ is the change in conductivity, $V_{SD}$ is the source-drain bias voltage, $\phi_{photon}$ is the photon flux, and $A$ is the area of the graphene channel. For sample A, the optical gain is as high as $1.7 \times 10^6$ when the power is 440 nW/cm$^2$ and $V_G = -55$ V. The high gain observed in our devices can be attributed to the high mobility of graphene, which yields multiple hole circulations in the graphene channel while the electrons remain trapped in the Chl*a* layer by the built-in field.[6] The large photoresponse characteristics observed in the hybrid graphene-chlorophyll devices suggests a feasible method of utilizing highly stable biomaterials for graphene photosensitization.

A comparison of different thicknesses of the Chl*a* film yields further insight into the photoresponse in our devices. In Figure 4a, we compare the $V_{SD}$ dependence of $\Delta I$ for sample C (~50 nm) and D (~10 nm). The characterization for the thickness of Chl*a* film is described in supporting information S2. The photocurrent is found to be linearly increasing with bias voltage for $V_{SD} < 1$ V. Based on Ohm's law $V_{SD} = I(L/W\sigma)$ and the expression for conductivity $\sigma = n_h e \mu_h$ (hole conduction), we found that $\Delta I$ varies linearly with $V_{SD}$ and that the slope is $\Delta I / V_{SD} = \Delta n_h e \mu_h (W/L) \propto \Delta n_h$. Therefore, the slope in the $\Delta I - V_{SD}$ curves represents the magnitude of doping due to the photogating effect. We found that the change in carrier concentration ($\Delta n_h$) enhances with increasing thickness, which is apparently originated from the stronger photogating effect in the thicker Chl*a* film. In Figure 4b, we show the power dependence of the responsivity for samples C and D, which also manifests a clear thickness dependence of the responsivity. Previous study on the interface between Chl*a* and metal revealed an exciton diffusion length of ~18 nm in the Chl*a* film.[44] Therefore, it is possible that the carriers can diffuse within the Chl*a* film and subsequently transfer to graphene, causing the thickness dependent photoresponse.

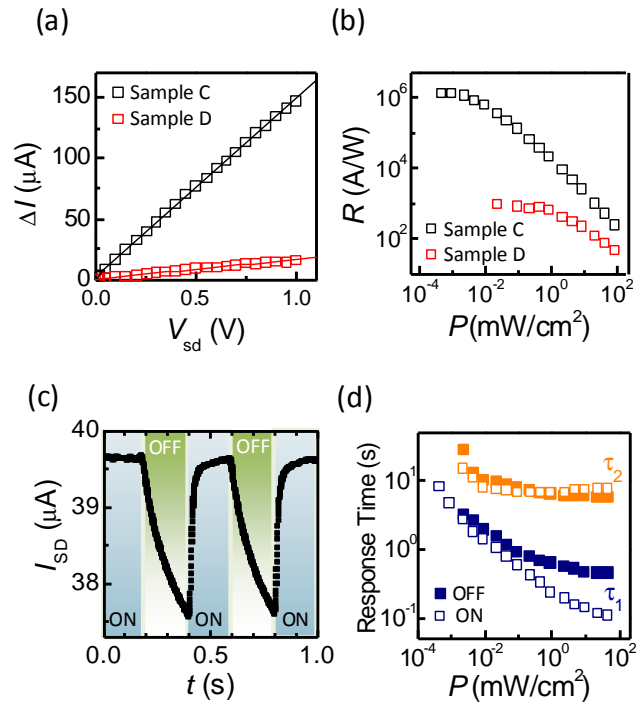

**Figure 4.** (a) Photocurrent as a function of source-drain voltage for samples C (50 nm) and D (10 nm). (b) Comparison of the responsivity for samples C and D. The saturated responsivity of samples C and D are $1.3 \times 10^6$ and $1.0 \times 10^3$ A/W, respectively. (c) Temporal photocurrent dynamics of sample A for $V_G = -55$ V and an excitation power of 22 mW/cm². Both on and off periods are 200 ms. (d) The excitation power dependence of the response time ($\tau_1 < \tau_2$) for light (open squares) and dark (close squares) periods.

Finally, we examined the photoresponse dynamics revealed by the time dependent measurement of the photocurrent. Figure 4c shows the variation of the channel current of sample A for a sequence of on and off periods of irradiation. In the hole conduction regime ($V_G = -55$ V), the device exhibits a positive gain in photocurrent with photoexcitation on and a diminution after the illumination shuts off. Each time dependence of the photocurrent can be fitted with two relaxations (supporting information S4), and the power dependences of the response time ($\tau_1 < \tau_2$) are shown in Figure 4d. We found that $\tau_2$ for on and off follows a similar trend that remains near 10 s for all power range. On the contrary, $\tau_1$ exhibits a stronger decaying tendency as the excitation power rises. Because $\Delta I$ is associated with the carrier density change ($\Delta n$) due to photogating, the $\Delta I - t$ curves represent the time dependence of charging and discharging in the Chl*a* film. Based on the time scale of the relaxation, it is possible that $\tau_1$ is related to the photogating effect from the Chl*a* molecules in close vicinity of graphene which leads to the instant photocurrent change. We observe that $\tau_1$ decreases with increasing excitation power. This behavior suggests that the transient photocurrent change is related to an instant surge in the hole carrier concentration and resulting charge transfer, which are more intense under higher illumination power. We note that the observed response time of the graphene-Chl*a* device ($\tau_{1,ON} = 0.11$ s and $\tau_{1,OFF} = 0.45$ s for P=44 mW/cm$^2$) is comparable to those in the graphene-PbS QDs phototransistors,[17] suggesting a critical role of the graphene-photosensitizer interfaces with organic Chl*a* molecules or ligand on PbS QDs. Furthermore, the response time $\tau_2$ may be correlated with the carrier diffusion and/or trapping in stacks of Chl*a* molecules, which is dominated by the bulk Chl*a* properties, such as mobility and trap density.[17]

## 4. Conclusion

We have demonstrated the prominent photoresponse and the highly back-gate-tunable photocurrent in the hybrid graphene-chlorophyll phototransistors. The devices exhibit a high gain of $10^6$ electrons per photon and a high responsivity of $10^6$ A/W, which can be attributed to the integration of high-mobility graphene and the light-absorbing chlorophyll molecules. The charge transfer at the interface and the effective photogating effect in the chlorophyll layer can account for the observed large photoresponse, which is confirmed by the thickness and time dependent studies of the photoresponse. The demonstration of the graphene-chlorophyll phototransistors with high gain envisions a feasible method to employ biomaterials for future graphene-based optoelectronics.

**Acknowledgments**

W.-H.W. thanks Juen-Kai Wang for the insightful discussions and Pin-Hao Sher for technical assistance. This work was supported by the National Science Council of Taiwan under contract numbers NSC 101-2112-M-001-020-MY2, NSC 101-2113-M-002-016-MY2, and NSC 100-2119-M-002-020.

**Supporting Information Available.** Device fabrication of graphene-chlorophyll phototransistors, chlorophyll-a functionalization and characterization, power dependence of graphene transport characteristics, and analysis of time-dependent photoresponse.

Supporting Information for

# Biologically inspired graphene-chlorophyll phototransistors with high gain

Shao-Yu Chen[1], Yi-Ying Lu[1,2], Fu-Yu Shih[1,3], Po-Hsun Ho[4], Yang-Fang Chen[3], Chun-Wei Chen[4], Yit-Tsong Chen[1,2], and Wei-Hua Wang[1*]

[1]Institute of Atomic and Molecular Sciences, Academia Sinica, Taipei 106, Taiwan

[2]Department of Chemistry, National Taiwan University, Taipei 106, Taiwan

[3]Department of Physics, National Taiwan University, Taipei 106, Taiwan

[4]Department of Materials Science and Engineering, National Taiwan University, Taipei 106, Taiwan

[*]Corresponding Author.

Tel: +886-2-2366-8208, Fax: +886-2-2362-0200. E-mail address: wwang@sinica.edu.tw (W.-H. Wang)

## S1. Device fabrication of graphene-chlorophyll phototransistors

Monolayer graphene sheets were mechanically exfoliated on the OTS-functionalized $SiO_2$ (300 nm)/Si substrate. The OTS-functionalized $SiO_2$/Si substrates can effectively reduce charged impurity scattering due to adsorbed molecules and the phonon effect, which is coupled to the substrates. Consequently, graphene on OTS-functionalized $SiO_2$/Si substrates exhibits high mobility, which is important for the demonstration of the large photoresponse. We applied both optical contrast and Raman spectroscopy to identify the single-layer graphene (SLG) sheets. We adopted the resist-free fabrication technique to avoid undesirable contamination through the lithography processes. After alignment of the SLG sheets and the shadow mask (commercial TEM grid) under the optical microscope (OM), we deposited electrodes with Ti/Au (5 nm/55 nm) by using an electron-beam evaporator at a base pressure of approximately $1.0 \times 10^{-7}$ torr. The channel length of the graphene devices is 12 $\mu$m.

After metallization, the graphene device was subsequently transferred into a cryostat (Janis Research Company, ST-500) for the initial characterization. The pressure in the cryostat is $2.0 \times 10^{-7}$ torr. Before characterization, we performed local laser annealing by focusing the laser (532 nm, 10 mW, spot size: 1 $\mu$m) adjacent to the device for 2 hours to remove the surface adsorbates, such as $H_2O$ and $O_2$ molecules in ambient conditions. The laser intensity on the sample is estimated to be 2.4 mW/$\mu$m$^2$. Local laser annealing is preferred over thermal heating, which may introduce cross contaminations from other part of the samples.

The electrical measurements were performed by using Keithley 237 source-measurement unit for the DC resistance measurement ($V_{SD} = 100$ mV for all measurements) and Keithley 2400 for back gate voltage. In the photoresponse measurements, we chose the semiconductor diode laser with a wavelength of 683 nm as a coherent light source. We note that the excitation wavelength for all photoresponse measurements is chosen to be approximately the peak value in the photocurrent action spectrum to maximize the photoresponse of the samples. The laser beam was guided into the microscope and focused by a 10X objective and the spot size is ~10 $\mu$m. Moreover, to acquire more information on the graphene/Chl*a* phototransistor *in situ*, micro Raman spectroscopy and micro PL measurements are integrated into the transport measurement system. The schematic of our measurement system is shown in Figure S1.

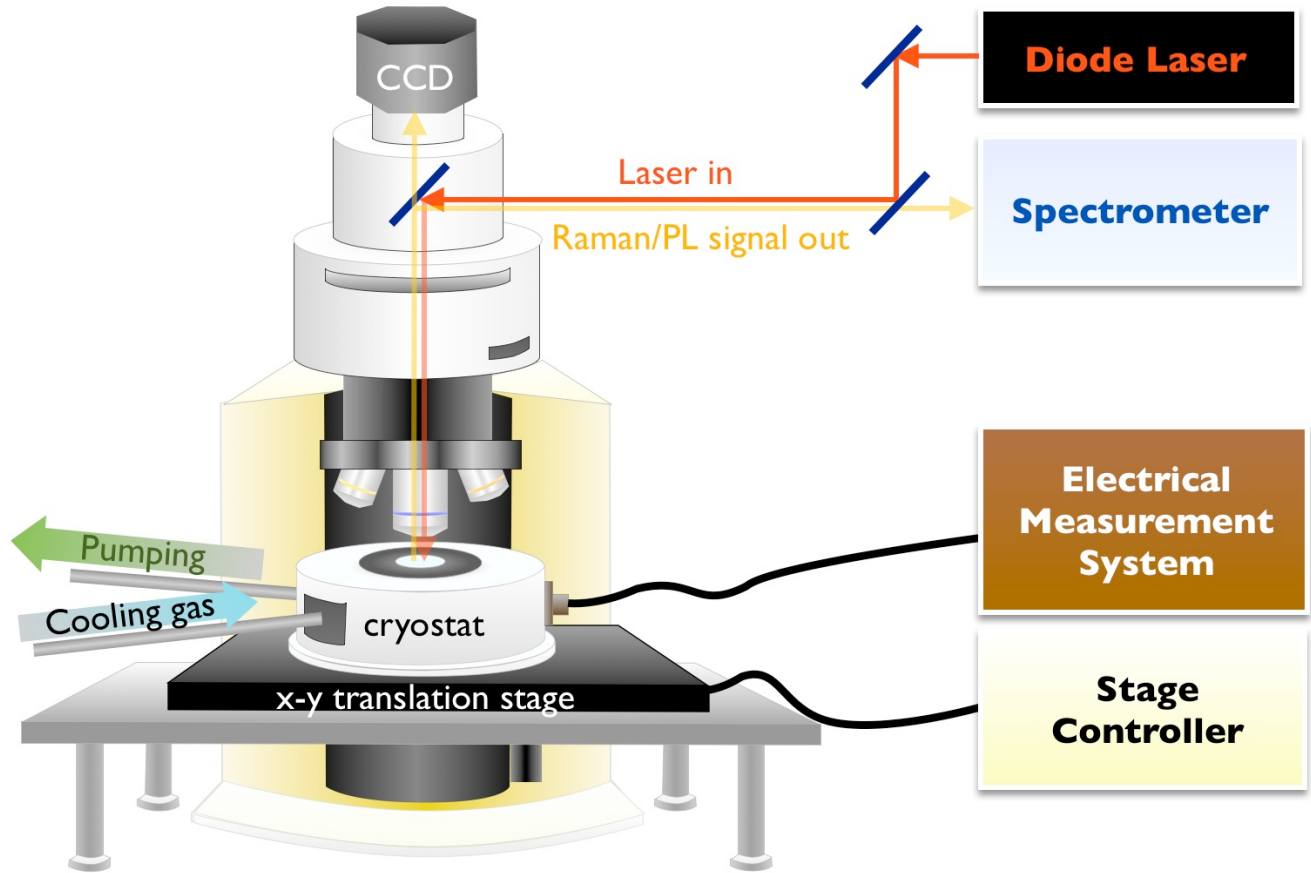

Figure S1. A schematic of the integrated cryostat/OM/μ-PL/μ-Raman/photocurrent measurement system.

## S2. Chlorophyll-*a* modification and thickness characterization

We chose Chl*a* because it is a light-absorbing material that exhibits good absorption in the visible-light spectrum for the graphene-based phototransistor. We applied commercial Chl*a* powder (Sigma Aldrich, 99.9%) and stored it in the refrigerator at $-20°C$. We chose ether as a solvent to prepare the Chl*a* solution of 1.5 and 50 $\mu$M following the functionalization procedure. In this article, we used the drop-casting method to functionalize the graphene devices with Chl*a* molecules. The thickness of Chl*a* layer can be determined by AFM with specific calibration procedure.

To determine the thickness of the Chl*a* film, we developed the following procedure. First, we chose a few layer graphene prepared by mechanical exfoliation. Before the Chl*a* functionalization, we measured the step-height of few layer graphene by atomic force microscopy (AFM) (Figure S2a). Next, we fabricated Au electrodes (60 nm), followed by drop-casting the Chl*a* solution. It is noted that the Au electrode of this test sample can be easily removed because of the lack of titanium, which would serve as an adhesion layer. After removing the Au electrodes, we obtain a clean silica surface, which is the same reference level before Chl*a* functionalization. Then, we measured the height of the Chl*a*/graphene bilayer (Figure S2b). The thickness of the Chl*a* layer can be determined by subtracting the thickness of the Chl*a*/graphene bilayer from the thickness of the pristine few layer graphene. The thickness of sample A in the main text is characterized as 40 nm. Notably, we obtained a surface roughness of the Chl*a* surface (0.37 nm) that is similar to the surface roughness of graphene (0.35 nm), which indicates a uniform functionalization.

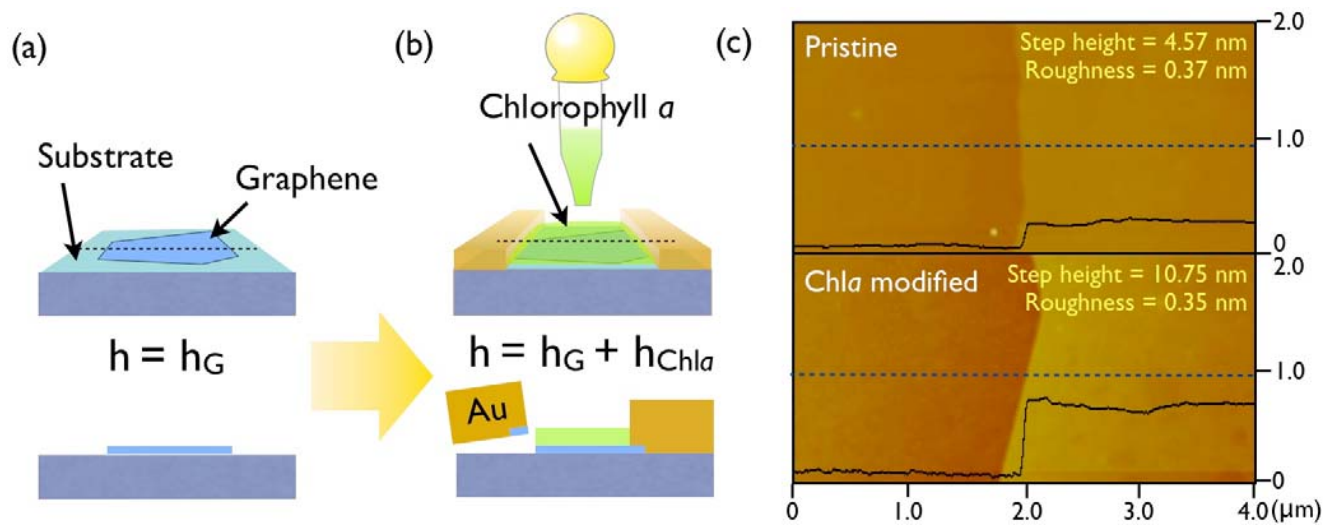

Figure S2. The schematics of the procedure to measure the thickness of the (a) few layer graphene and (b) the Chl*a*/graphene bilayer. (c) The AFM images of the few layer graphene (upper panel) and the Chl*a*/graphene bilayer (lower panel). The step height and roughness before and after Chl*a* functionalization are shown.

## S3. Power dependence of graphene transport characteristics

We showed that the minimum conductivity decreases as the excitation power increases (Figure 3d in the main text), which indicates a trend toward the intrinsic state of graphene. This behavior is consistently observed in our samples as shown in Figure S3 (upper panel), which may be attributed to the decreased charge inhomogeneity in graphene caused by the decrease in number of charged sites in the Chl*a* film. Moreover, we observed an increasing electron and hole mobility with increasing excitation power, as shown in Figure S3 (lower panel). The n-type doping due to the Chl*a* film results in a reduction in mobility due to charged impurity scattering. The charge transfer in the device reduces the n-type doping effect, which leads to a revived electron and hole mobility. This trend is generally valid, although sample-to-sample variation also exists.

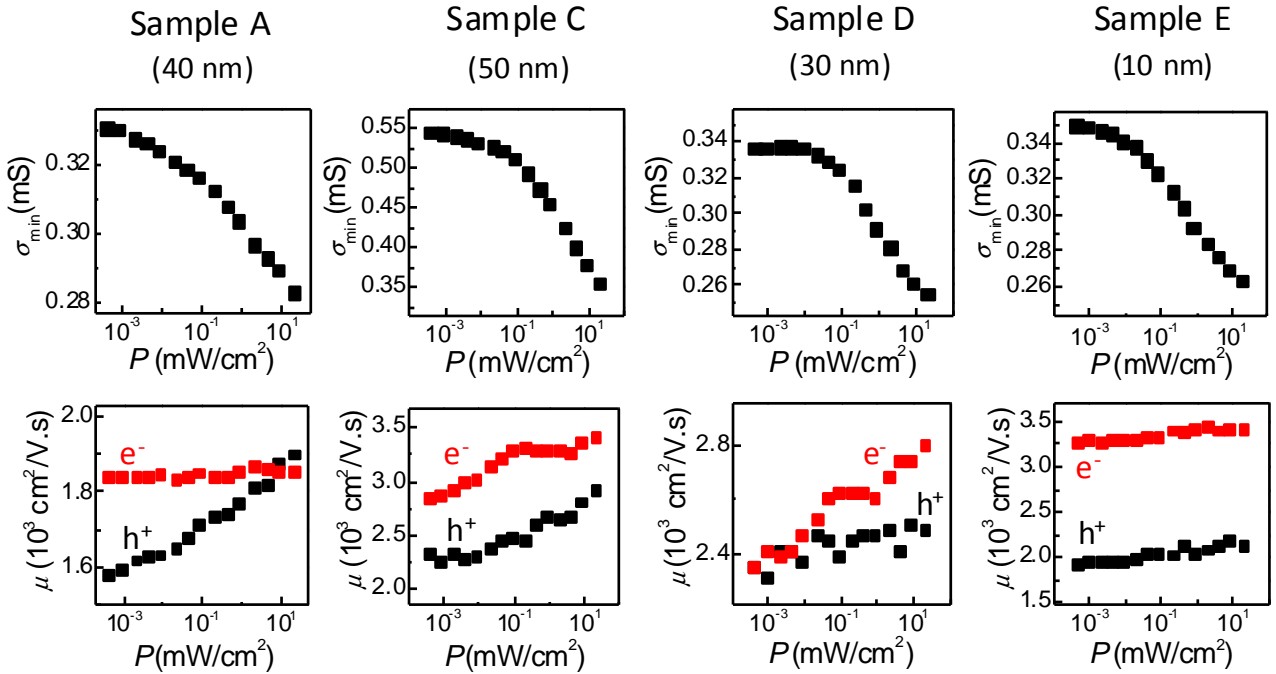

Figure S3. Minimum conductivity (upper panel) and electron/hole mobility (lower panel) as a function of excitation power for samples A, C, D, and E.

## S4. Analysis of time-dependent photoresponse

In this section, we present the dynamics of the photoresponse under different illumination powers and the fitting of the response time. In the high optical power region while the photoresponse is faster, all of the dynamics data were measured by the built-in sweep function of Keithley 237, which offers the fastest setting-reading cycle time as 1 ms. In our experiments, we adopted the setting-reading cycle of 24 ms to reduce the AC noise at 60 Hz. In Figure S4a, the time dependent photocurrent is recorded just after the shutter was turned on, which can be well fitted by two exponential functions. It is apparent that the faster photoresponse depends more strongly on the laser power than the slower photoresponse. Moreover, the photoresponse after the laser turns off (Figure S4b) exhibits similar behaviors. Figure S4c explicitly shows the turn-on and turn-off photoresponse for sample A under 10 nW illumination. We can reasonably extract $\tau_1 = 0.33$s and $\tau_2 = 6.85$s for the on time dependent photocurrent and $\tau_1 = 0.69$s and $\tau_2 = 6.73$s for the off time dependent photocurrent.

We estimate the quantum efficiency (QE) for the graphene-Chl*a* device (sample A) based on the equation $\Delta n = \tau \times QE \times \phi_{photon}$, where $\Delta n$ is the carrier concentration transferred from chlorophyll to graphene under illumination; $\tau$ is the carrier lifetime; and $\phi_{photon}$ is the photon flux illuminated on the sample.[S1] Because the photoexcited carriers that induce the photogating effect are mainly transferred at the fast response stage, we can estimate the QE by extracting $\Delta n$ at this initial stage. For illumination power of 10 pW, the photon flux ($\phi_{photon} = 1.5 \times 10^{13}$ $cm^{-2}s^{-1}$) induces an increase in charge accumulation density of $\Delta n = 1.2 \times 10^9 \, cm^{-2}$ during $\tau = 1$ second, which yields the QE of 0.008 %. Figure S4d shows the

QE under various illumination power. We observe that the QE decreases fast when the power exceeds 10 nW, which may be due to the saturation of the photogating effect.

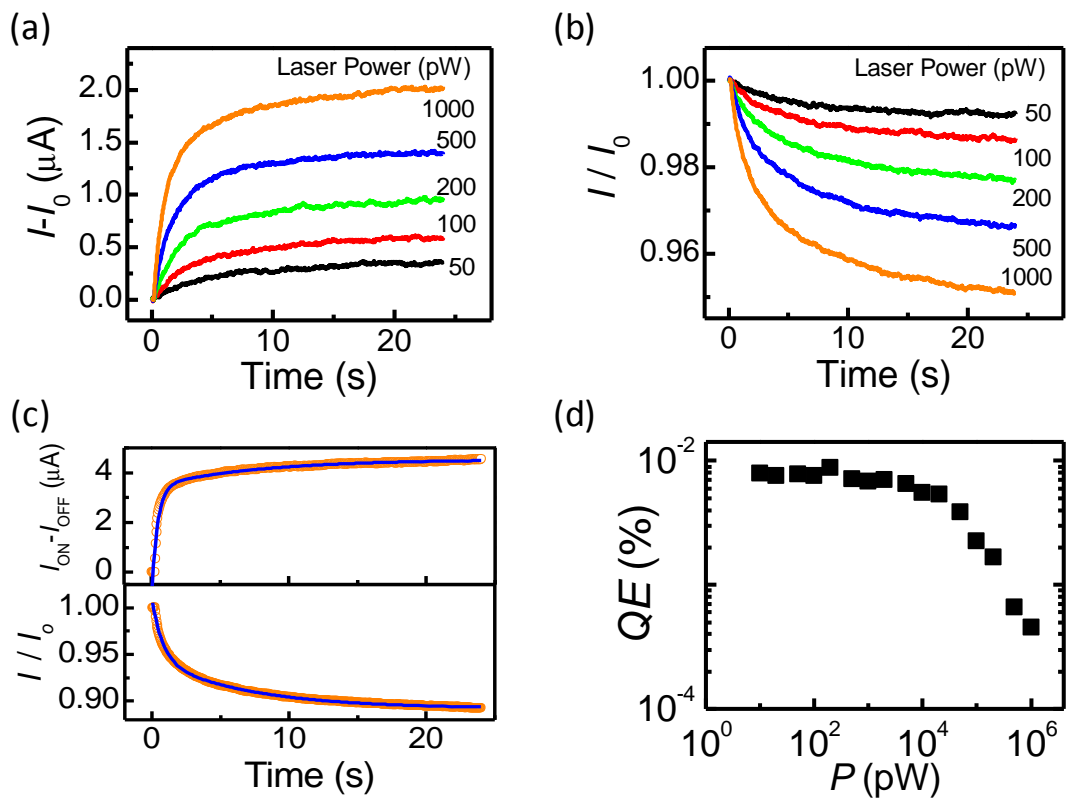

Figure S4. The (a) on and (b) off photoresponse dynamics are shown in various illumination powers from 50 pW to 1 nW. (c) The photoresponse are well fitted by two exponential functions for both on and off photoresponses under 10 nW illumination. (d) The quantum efficiency under various power.